# On an Approach to the Design of a Logical Model of Innovation Project Data

**L. A. Mylnikov[a] and A. V. Trusov[b]**

[a] *Perm State Technical University*
[b] *Perm Scientific and Technical Information Center, Federal State Administration, Russian Energy Agency, Russian Energy Ministry*
e-mail: leonid@pstu.ru, tav@permcnti.ru

**Abstract**—Questions concerning the development of a logical model of innovation project data, as well as those concerning the design of information systems for decision-making support in the management of innovation projects, are discussed.

**Keywords:** management, a model, decision-making support, information system, design, innovation
**DOI:** 10.3103/S0147688211030142

## INTRODUCTION

Innovation projects are dealt with in a variety of areas that are studied separately (technical, technological, organizational, economic ones, knowledge management) and employ various management mechanisms that are typical of these systems. This is why the theory and practice of innovation management were previously developed by solving local problems. These detailed problems generated many methods and approaches that solve minor specific problems. Because managing innovations as an entire system poses a problem, it is proposed that local problems should be solved as a subsystem of an innovation project, as the same innovation type, or as the same scientific and technical, organizational, technological, etc. idea. There are so many of these local problems that choosing and verifying the developed approaches becomes a whole new problem, which allows us to work out what types of innovation can exist and what challenges one faces when choosing ways of realizing them.

No management is possible without first collecting necessary information and presenting it in a suitable form. Presently, managing any information-based process (knowledge of a subject field) is inconceivable without information technologies.

Automation of innovation project management is problematic, due to the fact that no software systems exist that could cover all of the basic needs of the given area (Fig. 1).

The development of an appropriate information system is of iterative character depending on the type of problem in question. Viewing management systems primarily as information-based ones, many common features can be found (i.e., necessary information is supposed to be stored somewhere). Regardless of the used method of decision-making support (methods based on processing accurate practical data and presenting them in a form suitable for a person in charge of decision-making, or methods that process data from expert evaluations and polls), all of them use a limited range of data-storage methods. The differences lie merely in the logical model used in a concrete data model.

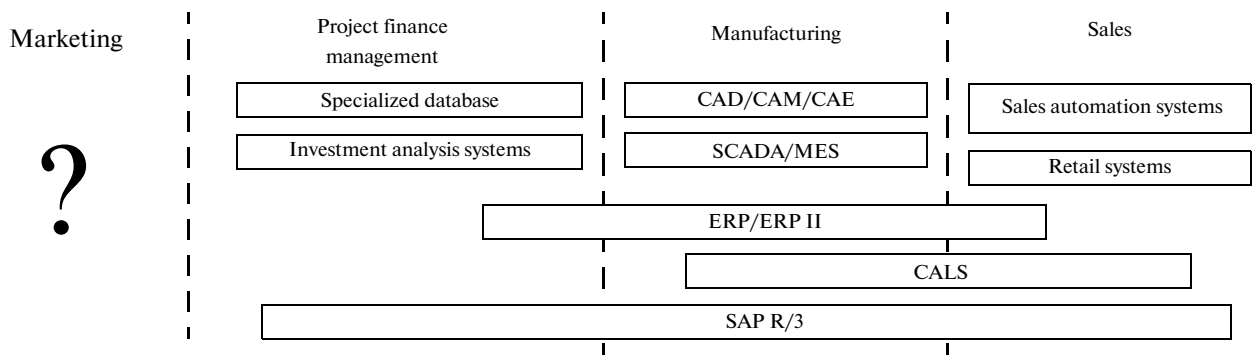

**Fig. 1.** Software-based support of the functional stages of an innovation project.

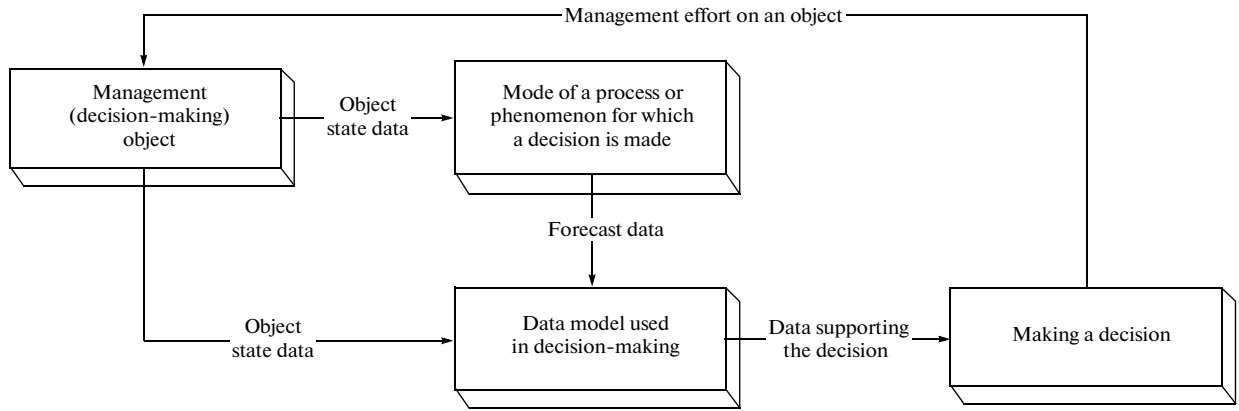

**Fig. 2.** Decision-making scheme using modern management methods.

## LOGICAL DATA MODEL

Decision-making theory is underpinned by the assumption that management can be adequate, provided that two data sources are combined: a model and data obtained in a management object [1]. Thus, adequate management of an innovation project should be combined with model management and data management (Fig. 2).

For this purpose, a modular data model needs to be developed that could make use of the experience from applying already existing methods and information presentation. Using both of these techniques is possible, provided that one succeeds in designing a hierarchical structure of goals and decisions similar to the one proposed by Thomas Saaty [2].

Decisions can be made at each stage of project realization. In order to understand a problem, a tree should be designed that is similar to the one that is used in the hierarchy analysis method (HAM) [2] with the decomposition process (splitting the criteria into constituent criteria) lasting as long as necessary, as in the HAM. The major distinction of the HAM that results in its being non-applicable to the management of innovation projects is that possible outcomes are not known. The solution thus cannot be reduced to choosing one of the variants.

When a project is performed, information on the resources at hand exists, as well as on the additional resources that can be used. The distinction of decision-making methods is that basically they do not help one to make a decision but assist in building a coherent decision-making structure and presenting the information in a suitable form for a person in charge of decision making. This approach is realized based on specific data structures that do the job of organizing and analyzing the available information. A data structure normally reflects a tree-like structure or a graph, provided that a subject field requires counterbracing on a relational data model and further detailing of the hierarchy given in Fig. 3.

In this approach, the leaves of the tree (see Fig. 3) represent the maximum possible detailing of a project. The corresponding tree-like structures can be considered as measurements (in the terminology of the systems of analytical data processing) and the numerical values that are compared to them as measures. The apparent simplicity is made complex by the fact that some decisions can be interconnected, i.e., can depend on each other (e.g., either this solution or another, or both). This fact is taken into account if the connections of the elements are viewed as the logical operators "AND" or "OR." We then obtain a morphological tree, which is well-known in the theory of design. Unfortunately, these interconnections can be more complex than logical operators. For example, for innovation projects, there are always some restrictions that can be made on each specific project. These are such data as the time of a project payback and expenditures spent or planned to be spent on realization, as well as those that can be determined using the development curve of an innovation project. The description of complex interconnections is used in expert systems of the production type (Fig. 4). During comparison of the rules to the connections of the tree-graph, complex interconnections can be considered in an innovation system.

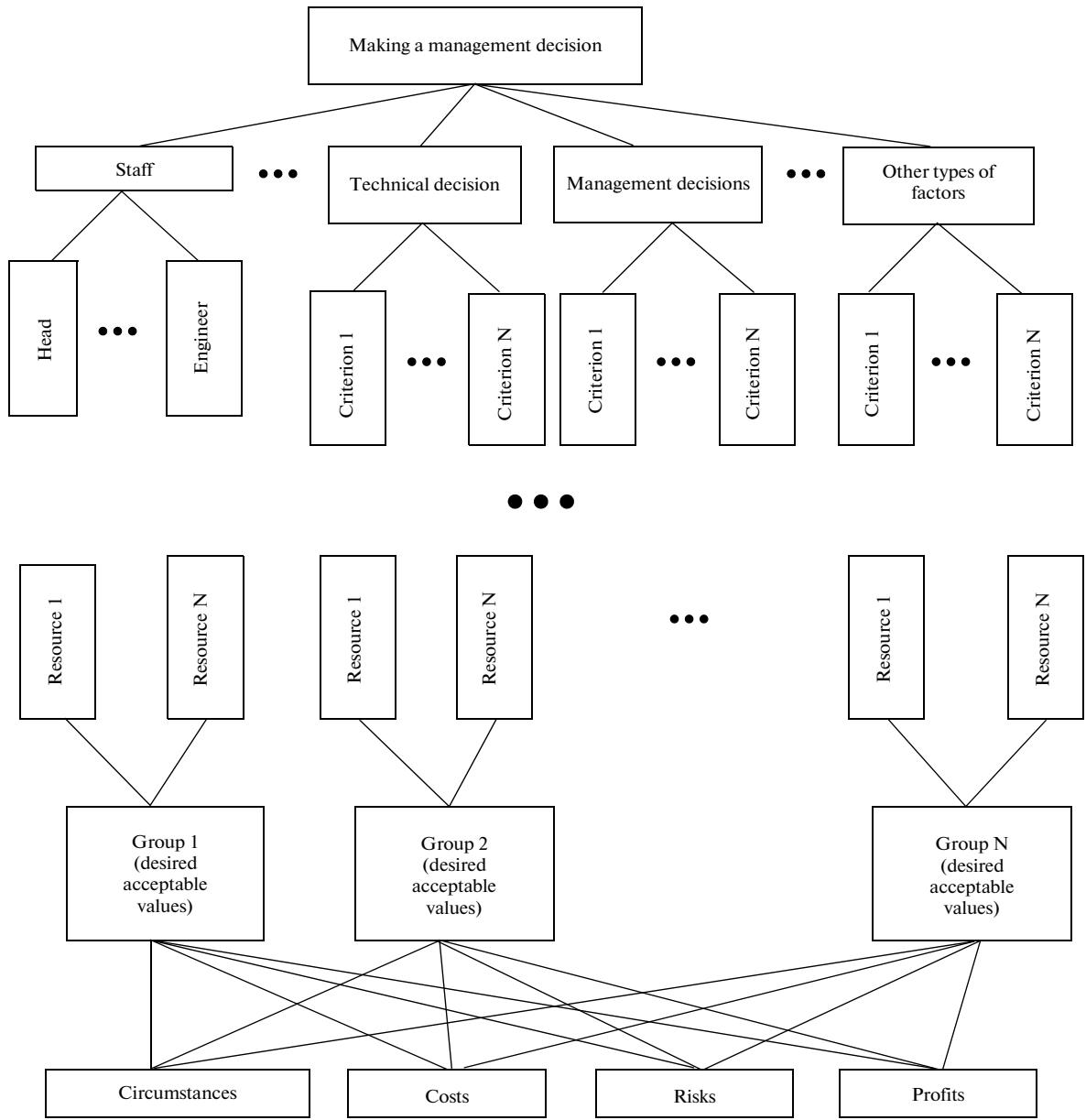

**Fig. 3.** Decision-making hierarchy in innovation project management.

It is known that logical rules are used successively in a production expert system, which fits the tree-like structure perfectly. This means that using this approach, most complex dependences based on logics can be built.

*The production rules* are interpreted using the structure IF $\alpha_i$, THEN $\alpha_j$. We may have the following production: IF $\alpha_1$ and $\alpha_2$ and $\alpha_3$ and $\alpha_4$, THEN $\alpha_5$.

Or, schematically:

$$\alpha_2 \rightarrow \begin{matrix} \alpha_4 \\ \alpha_3 \end{matrix} \rightarrow \alpha_5.$$

However, a simple logic is not enough to evaluate innovation projects. Management is performed by choosing from a limited number of some numerical characteristics. These may have different values and

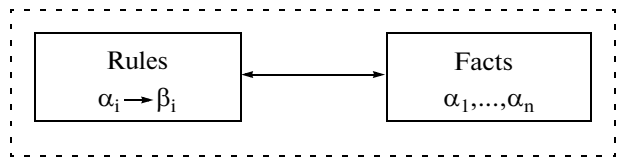

**Fig. 4.** Diagrammatic logics of the operation of an expert system for production.

204 MYLNIKOV, TRUSOV

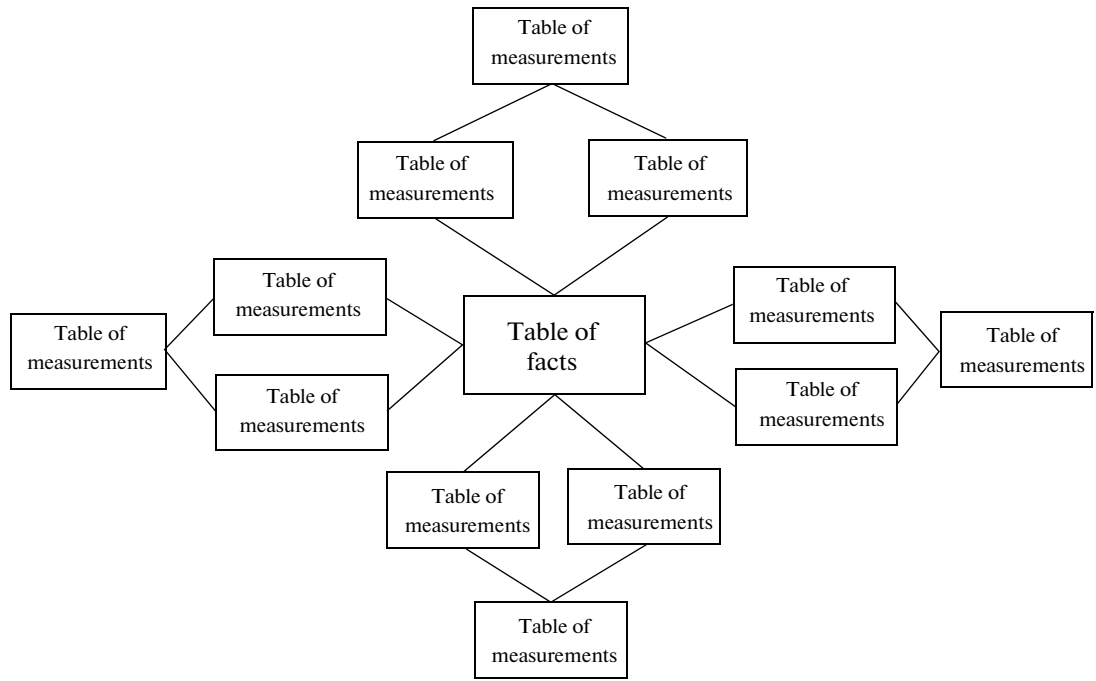

**Fig. 5.** Database scheme for a structured decision-making support based on a relational data model.

content depending on a leaf or a node on a graph to which they are compared.

Thus, to each element of our tree or graph, we compare some discrete table of characteristics or a function. The sets of characteristics compared to homogenous elements should be the same. With this achieved, evaluation criteria methods can be used and the structure becomes sufficient for analyzing and developing decisions, as it combines the elements of production and frame expert systems, systems of analytical data processing (OLAP), and systems based on the analysis of hierarchies, as well as methods for mathematical programming. All the above methods are known means of decision-making support, while the described system structure includes features of certain methods of decision-making support and allows one to combine them in arbitrary combinations depending on the problem at hand.

Realization of these functions using the traditional methods of multidimensional data storage can be difficult, as the traditional technology places requirements on the organization of data interconnections according to one of the known schemata (a snowflake schema or star schema); as well, it does not allow one to build mutually directed hierarchies with the same low elements shown in Fig. 3. This structure can be realized using two schemas of the star type, whose finite tables (leaves) will be common. In order to operate with these data structures, it is necessary to introduce a new operator that would choose the main fact table and thus to introduce a new data scheme to one of the existing star schemata or combine the fact tables and finally obtain the structure presented in Fig. 5.

As has been shown above, for decision-making and management, there should be a model or algorithm that would be able to operate the data. For this, one of the approaches employed in DataMining, viz., a decision tree, [3] can be used. This approach allows the user to operate the suggested data structure and perform logical functions, while computing operations that involve the model operation can be realized by a database program.

## 2. A DATA MODEL IN INFORMATION SYSTEMS

Traditionally, information and analytical systems based on logical data models are developed for the generation of analytical reports on various queries made to a subject field. Queries can be of two types:

• Standard — these queries are described at the stage of the system development and are further renewed only when data are loaded into a subsystem. Further, users can access these reports. They should require no reference to the relational or multidimensional base of the analytical subsystem. Reports should be in the XML dialect format.

• Special — these queries utilize the connection to the multidimensional base of the subsystem and enable fairly complex reports, despite the fact that they require a highly trained analyst and connection to a local network where there is a server.

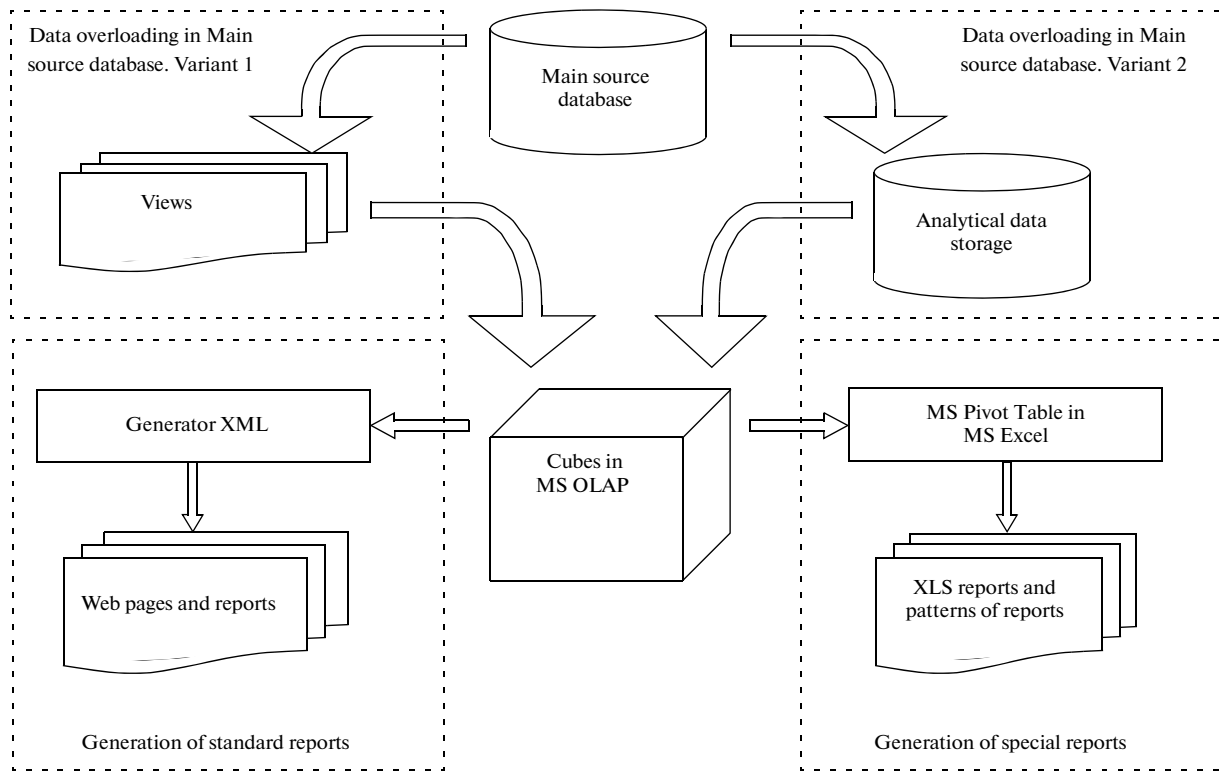

**Fig. 6.** Subsystem structure scheme.

Any reports are generated using OLAP technologies and a multidimensional base of the subsystem. References to the relational storage are acceptable but not desirable. One possible architecture of such a subsystem can have the form shown in Fig. 6.

This system can create two types of reports:

• Static – these are realized as queries to the multidimensional base and stored in the XML format. The accurate list of these reports should be verified at the realization stage. The total number of such reports should not exceed ten. Reports can be viewed in MS IE or in other environments. Accompanied by XSL descriptions, reports can be easily published in the Web.

• Dynamic – these are realized in Excel patterns using PivotTables. Based on these patterns, analysts will be able to obtain necessary reports by ordinary manipulations. These patterns can be accurately specified at the realization stage, but their total number should not exceed 15 (5 patterns per each of the cubes). Microsoft Excel is used to view and prepare dynamic reports.

Any rules and algorithms for developing management efforts can be employed in the generation of reports. The data structure and methods for making management decisions are thus interconnected.

Further, application of the model based on the data and analytical processing facilitates "user–algorithm" interaction by utilizing visual human–machine interfaces (Fig. 7).

Problems of industrial engineering emerge in the transition to innovation projects (e.g., management and manufacture planning).

Frederick Taylor and Henry Gantt are credited with the development of the methods of enterprise management in the early 20th century [4–6]. They were responsible for creating industrial planning as a discipline. The works by Frederick Taylor and Henry Gantt laid the foundation for scientific disciplines that emerged in the mid-20th century, such as Industrial Engineering, engaging in the management and organization of manufacturing, as well as Operations Research [7]. In the early 1960s, the effort towards automatic Inventory Control got underway in the USA [8, 9]. Due to the active growth of large-scale and massive manufacturing of consumer goods and trade after the Second World War, it became obvious that using mathematical models of demand and inventory control planning can save expenditures frozen as supplies and incomplete manufacturing. The late 1960s were marked by the works by Oliver Wight, who suggested regarding industrial, supply, and marketing departments all together during the automation of industrial enterprises. In the publications by Oliver Wight and the American Production and Inventory Control Society that are dedicated to inventory control and manufacturing management, planning algorithms were formulated that are now known as MRP (Material Requirements Planning), which emerged in the late 1960s [10], and MRP II (Manufacturing Resource Planning), which was used in the late 1970s until the early 1980s [11]. The concept of computer integrated manufacturing (CIM) came about in the early 1980s and involves the integration of flexible manufacturing and its management systems as management and planning systems (for which ERP and MRP II are used). The CALS methods (Computer-aided Acquisition and Logistics Support) appeared in the 1980s in the US Department of Defense in an attempt to boost management and planning efficiency for the ordering, development, manufacturing organization, delivery, and operation of military equipment. CALS leads to one-time data input, their storage in standard formats, interface standardization, and digital exchange of information between all the organizations and all their departments that participate in a project [12]. These methods have proven to be highly efficient. Some aspects of CALS are now standardized in the ISO international organization.

Integrating the tasks of managing and planning in the early 2000s brought about a new manufacturing trend in the USA. This was known as "Supply Change Optimization" (now known as "Supply Chain Management").

The model data thus becomes the basis for a universal information system of an enterprise bringing together software-based support of the functional stages of an innovation project (see Fig. 1); this enables integration of all the processes that make up a closed system of innovation development of an organization within a universal information space:

• Forecasting (determining the moment when a new item should be produced, having a good knowledge of current marketing trends, and estimating the lifespan of technical and technological decisions).

• Production of new products using cutting-edge scientific technologies.

• Management and planning with regard to innovative production processes.

## CONCLUSIONS

The authors of the present paper have developed an information and analytical system that, if introduced, can cut the costs that are incurred in project analyses that are funded by the Perm region authorities. It could also help reduce the amount of time spent on project analysis at the stage of its design. The suggested formalization provides an opportunity to preserve the operability of the regional system of project selection, despite numerous attempts to update the administrative management system of the region and the replacement of the core personnel.

## ACKNOWLEDGMENTS

The authors are grateful to the Perm Region Government for authorizing this research and subsequent development of the regional information analytical system of innovation support InnoNet.